\documentclass{article}

\usepackage{arxiv}

\usepackage[utf8]{inputenc} 
\usepackage[T1]{fontenc}    
\usepackage{hyperref}       
\usepackage{url}            
\usepackage{booktabs}       
\usepackage{amsfonts}       
\usepackage{nicefrac}       
\usepackage{microtype}      
\usepackage{lipsum}
\usepackage{graphicx}
\graphicspath{ {./images/} }

\usepackage{graphicx}
\usepackage{tikz} 
\usepackage{amsmath}
\usepackage{rotating, graphicx}

\newcommand{\bbeta}{\mbox{\boldmath $\beta$}}
\newcommand{\bpsi}{\mbox{\boldmath $\psi$}}

\newcommand\indep{\protect\mathpalette{\protect\independenT}{\perp}}
\def\independenT#1#2{\mathrel{\rlap{$#1#2$}\mkern2mu{#1#2}}}

\title{Trial emulation and survival analysis for disease incidence registers: a case study on the causal effect of pre-emptive kidney transplantation}

\author{Camila Olarte Parra \\
    Department of Applied Mathematics, \\ Computer Science and Statistics \\
    Ghent University \\
    Ghent, Belgium \\

   \And
 Ingeborg Waernbaum \\
  Department of Statistics\\
  Uppsala University\\
  Uppsala, Sweden \\
  
  \And
  
  Staffan Schön \\
  Swedish Renal Registry \\
  Jönköping County Hospital \\ 
  Jönköping, Swedem \\
  
  \And
 Els Goetghebeur \\
  Department of Applied Mathematics, \\ Computer Science and Statistics \\
    Ghent University \\
    Ghent, Belgium \\
}

\begin{document}
\maketitle
\begin{abstract}
Numerous tutorials and research papers focus on methods in either survival analysis or causal inference, leaving common complications in medical studies unaddressed. In practice one must handle problems jointly, without the luxury of ignoring essential features of the data structure. In this paper, we follow incident cases of end-stage renal disease and examine the effect on all-cause mortality of starting treatment with transplant, so-called pre-emptive kidney transplantation, versus dialysis. The question is relatively simple: which treatment start is expected to bring the best survival for a target population? To address the question, we emulate a target trial drawing on the Swedish Renal Registry to estimate  a causal effect on survival curves. Aware of important challenges, we see how previous studies have selected patients into treatment groups based on events occurring post treatment initiation. Our study reveals the dramatic impact of resulting immortal time bias and other typical features of long term incident disease registries, including: missing or mismeasured covariates during (the early) phases of the register, varying risk profile of patients entering treatment groups over calendar time and changes in risk as care improves over the years. With characteristics of cases and versions of treatment evolving over time, informative censoring is introduced in unadjusted Kaplan-Meier curves and also their IPW version is no longer valid. Here we discuss feasible ways of handling these features and answer different research questions relying on the no unmeasured \textit{baseline} confounders assumption.
\end{abstract}


\section{Introduction}\label{Intro}

For decades now, the Randomized Clinical Trial (RCT) enjoyed the status of bringing gold standard evidence to inform clinical decisions \cite{Jones2015FateGoldStandard}. While advantages of this design are undeniable, the call for additional real world evidence sounds ever louder. This stems in part from the restricted and somewhat artificial setting of the randomized experiments, which challenges transportability of results to real world target populations \cite{McGinnis2010Redesigning,Dahabreh2020Transportability}. It is further the fruit of growing data resources of various types harbouring information from much broader natural target groups, which can now be mined using new developments in causal inference and beyond.  

Today's evidence supporting clinical decisions is thus also drawn from observed exposures, both within randomized trials, as requested in the ICH E9 appendix on estimands \cite{ICHE9Addendum}, and in the absence of trials. This becomes the primary evidence source when treatment cannot be randomized due to ethical or practical reasons as is typical for organ transplantation, for instance. A well designed (comprehensive) cohort of diseased patients may then bring the best chance of obtaining real world evidence. The latter should ideally be cast in clinically interpretable measures. Hence risk differences will be preferred over hazard ratios even though the latter may be an essential vehicle to arrive at the former. 

For a range of chronic diseases, population based incidence registers following patients from disease onset have been built and maintained over years. These are now important data sources for investigations of long term outcomes. Accrual over many calendar years also comes with additional challenges. Earlier entries are automatically subject to longer administrative  censoring times. When patient profiles and/or general level of care changes (improves) over calendar time several consequences must be addressed. First, assuming the study population is our target population, the population average survival curve, as estimated by Kaplan-Meier (KM) will be biased. Indeed, the longer administrative censoring times may then come with better survival chances. This informative censoring can be handled by simply adjusting for registry entry time. That well-known fact \cite{Booth2020Survivalovertime,Mader2020TimePattern}, gets easily forgotten given the robust reputation of the KM curve. It is not remedied by inverse probability of treatment weighting (IPW) adjustment which may address covariate imbalance across treatment groups. Second, having learned how entry time impacts survival on one or either treatment, the causal question of interest may shift from the full study population to what recent or even future patients can expect to benefit from their choice of treatment. Still, careful analysis of the available cohort will lay the foundation of such insight.         

When analyzing the effect of a point exposure on survival, one will obviously need to adjust for confounders associated with treatment at the time of treatment decision. It is then a great advantage that registries at the national level with broad coverage  typically have a well worked out protocol carefully defining  the set of patient characteristics to be included by all centers at the time of patient entry into the registry. Naturally this set may get updated after a number of years on a given date
to include additional covariates, responding to progressing insight in prognostic factors or more easy access to the (good) measurements.    

What happens during later follow-up tends to be much less controlled or harmonized as it emerges over long periods of time in a range of settings with more or less support for data measurement. Tight control would be extremely demanding at that level. It is hence important to understand what can be estimated when relying on a common set of baseline covariates without access to regular time-varying covariate measurements. 

As with clinical trials, the target estimand can either follow the intention-to-treat (ITT), per-protocol or as-treated principle addressing a corresponding causal question. In the randomized trial, the  ITT analysis commonly estimates the causal effect of being {\it assigned to a particular treatment} regardless of the adherence to it. In the observational setting it will also pertain to a point exposure which could be controlled at a specified time of `treatment' onset common to the available treatment options, such as treatment assigned, prescribed or initiated. Here too ITT marginalizes over subsequent treatment (intensity). An appreciation of exposure levels that follow in the study population  will deepen our understanding  of exposure and influence transportability of the estimand \cite{Dahabreh2020Transportability,Lee2020Generalizability,Vanderweele2013MultipleVersions}. Of course, before comparing outcomes of treatment groups conditional on covariates in the emulated trial, explicit adjustment for baseline confounders is required. This will ensure exchangeability before averaging over a chosen distribution of baseline covariates. This could be the covariate distribution observed in the full study population (average treatment effect, ATE) in the treated (ATT), the non-treated (ATNT) \cite{Goetghebeur2019} or any other relevant distribution.      

A per-protocol analysis targets the effect of adhering to a treatment regimen as established by the researcher. Strategies to deal with deviations from this regimen must then be specified. One approach restricts analysis to patients fully adhering to the treatment protocol or censors patients as they deviate from it. The latter may introduce informative censoring as time-varying factors likely influence both the treatment path and the outcome. Therefore, besides adjusting for baseline confounders, the per-protocol effect also requires adjusting for time-varying confounding. 

Finally the as-treated effect in RCTs pertains to treatment actually received (possibly for a given duration), rather than randomised to. No longer under the protection of randomisation, this approach typically involves accounting for both baseline and time-varying confounding, even in the context of an RCT \cite{CausalInference}. 

Through trial emulation, observational data can be used to mimic as closely as possible the data set-up that would have been aimed for in a target trial designed to answer the clinical question. It helps avoid bias frequently encountered in observational studies, e.g when allowing patient eligibility to rely on information obtained after treatment onset \cite{Hernan2016BigData, Hernan2016TargetTrial}. 

In this paper, we present a case-study where trial emulation draws on the Swedish Renal Registry, a nationwide research register. The research question investigates the total effect of immediate kidney transplantation versus starting on dialysis on all-cause mortality. The complications encountered and approaches taken apply quite generally to long term disease registers beyond nephrology. In Sweden and the Nordic countries, research on long term effects of a variety of chronic diseases is conducted through linkage of incidence registers and administrative registers with individual level data, e.g., the in-hospital register held by the National Board of Health and Welfare. The resulting data sets constitute a high-quality observational data resource for researchers used to both support and generate new hypotheses for a wide range of diseases.

In what follows we introduce our case study in more detail in Section \ref{Case-study}, we elaborate on the targeted estimands in Section \ref{Estimands} to discuss the estimation approach for ITT and an as treated analysis allowing for non-random treatment switch while relying on a sufficient set of baseline covariates for non-informative censoring, as described in Sections \ref{Potential_survival} and \ref{Methods_AFT}. The results derived from our case-study are presented in Section \ref{Results}, followed by a note on the existing software packages to aid these analyses in Section \ref{Software} and we end with a discussion on strengths and weaknesses of the approach taken and results obtained, relative to what is currently in the literature in Section \ref{Discussion}.  

\section{Case-study}\label{Case-study}

As kidneys are vital organs, patients reaching end-stage renal disease (ESRD) need treatment to survive. The two main alternatives are dialysis or kidney transplantation, collectively known as renal replacement therapy (RRT). Several studies consider how the modality of RRT impacts survival. Specifically, one aims to determine whether and by how much {\it patients} with immediate transplant, so-called pre-emptive kidney transplantation (PKT), have better {\it survival} than they would have from starting with dialysis, possibly followed by delayed transplant \cite{Abramowicz2016SR}. 

\begin{figure}[hbt!]
    \centering
    \begin{tikzpicture}
    \node (v0) at (4.00,2.50) {Death};
    \node (v1) at (-4.00,2.50) {Immediate transplantation (PKT)};
    \node (v2) at (0.00,5.00) {Confounders, $\pmb{Z}$};
    \node (v3) at (0.00,0.00) {Delayed transplantation};
    \draw [->] (v2) edge (v0);
    \draw [->] (v2) edge (v1);
    \draw [->] (v1) edge (v3);
    \draw [->] (v3) edge (v0);
    \draw [->] (v1) edge (v0);
    \end{tikzpicture}
    \caption{Causal diagram for the effect of immediate vs delayed transplantation on death}
    \label{fig:DAG}
\end{figure}
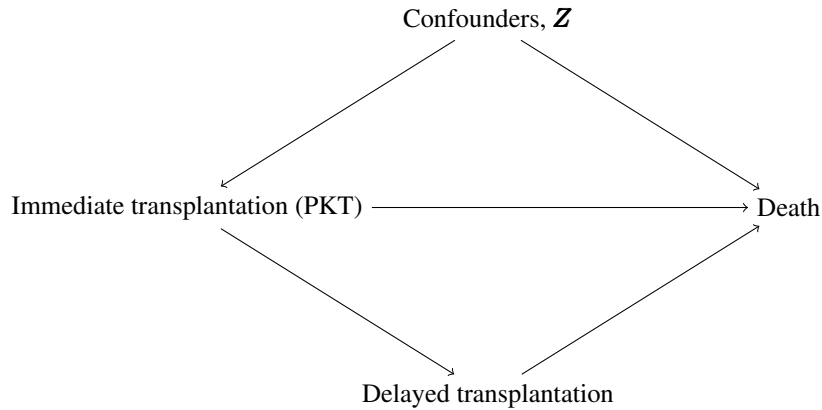

A systematic review of this research question \cite{Olarte2018Protocol} identified  published studies, most of them suffering from  avoidable biases. Those who worked with transplant registers are limited to RRT patients receiving a transplant and obtained retrospective information on when they reached ESRD and started dialysis. The restriction to patients living long enough to obtain the transplant results in immortal time bias \cite{Berthoux1996,Goldfarb2005Duration}. To account for this, some condition on the amount of time spent previously on dialysis lacking correction for truncation. This analysis ignores the mechanism of selecting subjects who 1) started RRT with dialysis in response to covariates then available and 2) have survived long enough to undergo transplantation \cite{ Haller2017Austria,Witczak2009Norway,Milton2008Australia,Kasiske2002US}.
Conditioning on covariates measured at the time of transplant or beyond (graft function or graft rejection) amounts to adjusting for events on the causal path from treatment initiation at RRT to survival \cite{Amaral2016, Cransberg2006}, another approach  well known to introduce bias (Figure \ref{fig:DAG}) \cite{Hernan2004SelectionBias}.

Analyses typically start from hazards and hazard ratios whose estimation entails limited additional modeling assumptions when censoring is non-informative or explainable \cite{Hernan2010Hazards}. Semi-parametric (extended) proportional hazards models are fast and stable to implement with residual plots to help assess model assumptions. Estimation of these building blocks avoids modelling the nuissance study-specific censoring mechanism per se. Derived summary measures, such as Z-specific or Z-standardized survival curves, carry direct clinical interpretation.
When the models are oversimplified however (sometimes when assuming time constant hazard ratios over the long haul) or the hazards are naively interpreted as causal contrasts for populations of survivors at time $t$ since RRT, problems arise. 

In what follows we overcome the  problems through an alternative design, transparent analysis under clearly stated assumptions and outcome parameters chosen for clinical interpretability. An incident RRT registry allows us to  mimic the ideal study which \textit{randomises} patients at RRT onset over either PKT or dialysis first. These exchangeable groups in terms of measured and unmeasured baseline prognostic variables are followed until death or administrative censoring. 
The nationwide Swedish Renal Registry (SRR) is such a cohort with carefully collected data since 1991 from all RRT units in Sweden, including 100\% of transplanted patients and at least 95\% of dialysis patients \cite{Schon2004Sweden,SRRReport}. Today, it records the following covariates when patients enter the register: date of RRT diagnosis, demographic variables at RRT onset, cause of kidney disease, RRT modality, comorbidities, kidney function and survival status \cite{Mollsten2010SRR}. Some covariates are introduced into the registry only years after it started: comorbidities (diabetes, hypertension, ischaemic heart disease, cerebrovascular disease and peripheral artery disease) since 1998 and kidney function since 2008. 

A sufficient set of measured baseline confounders justifies the assumption of ‘no unmeasured baseline confounding’ (NUBC) \cite{CausalInference}. Allowing various estimation strategies to account for differences between observed treatment groups in baseline characteristics, prognosis and potential benefit from treatments. When key covariates enter the registry late, one must either limit the analyses to the period where they are available, or consider imputation tyipically assuming missingness at random. With the sufficient set on board, causal effect analyses strategies include outcome regression, stratification, matching with or without propensity score, IPW or a combination in so-called doubly robust estimators. As we describe below, simple propensity score methods may not be valid in this context.

A well chosen contrast between Z-specific or Z-standardised average survival curves for the treatments considered  represents the specific population average causal effect under the NUBC assumption which we formalize in section \ref{Confounder}. An ITT analysis estimates the \textit{total} effect of treatment assignment, comparing \textit{marginal} survival curves between arms. This effect measure naturally averages over subsequent treatments: some patients starting on dialysis may get transplanted later, while others may die or be censored before ever getting a transplant. To interpret the effect of starting with dialysis, and especially with an intention of transporting results to new populations, one will wish to acknowledge the distribution of time to transplant among the dialysis starters. This is likely dependent on country-specific organ availability and transplant policies.   
Also, patients on either arm who receive a transplant may experience a graft failure requiring a new kidney or going on dialysis. The per-protocol and as-treated analyses must allow for non-random switching off the original treatment modality. This typically involves relying on time-varying covariates which we lack in the nationwide SRR. In sections \ref{Estimands} and \ref{Methods_AFT} we involve accelerated failure time models which allow for such non-random switch while relying on the no-unmeasured {\it baseline covariates} assumption for estimation.
  
Our set-up reminds of the work of Danaei et al.\cite{Danaei2013StatinsEmulation} who compare the effect on survival of initiating statins for the primary prevention of coronary disease versus not (yet). As in our case, those who do not start the treatment of interest (statins or PKT) at time $t$, may start later, and the ITT analysis averages over such changes in actual exposure as they naturally occur in the studied population. The `treatment’ comparison is then one of immediate vs delayed treatment initiation, where the latter comes as a compound treatment of dialysis possibly followed by transplant. This idea of a compound treatment can be seen in other settings too, like in oncology when the interest is to compare initial chemotherapy to reduce the tumour size prior to surgery, with surgery without delay \cite{Vergote2010Oncology}.

Mindful of the above considerations, we next define our estimands of interest in more detail before engaging in a well motivated protocol for analysis.

\section{Causal estimands of interest} \label{Estimands}

We aim to estimate the causal effect on mortality of starting RRT treatment with PKT rather than dialysis in a population eligible to receive either. Our outcome of interest, $T$, is time from RRT onset to death. Using the potential outcomes framework \cite{Rubin1974CausalEffect}, we consider the potential survival time from RRT onwards under two alternative possible exposures: $T_1,$ when a kidney is received without previous dialysis and $T_0,$ when RRT starts with dialysis. For this to make sense, we constrain the population to those patients for which both exposures are possible in principle, thus satisfying the positivity assumption \cite{Petersen2012Positivity}. To determine this (sub)population both statistical and clinical arguments enter, as discussed in section \ref{Method_Positivity} below.

In RCTs, an \textbf{ITT analysis} would typically estimate the total effect on time to death by comparing the survival distribution following PKT assignment, $S_1 (t)= P(T_1>t)$, with the survival distribution following dialysis assignment at RRT onset, $S_0 (t)= P(T_0>t)$, ignoring whether a delayed transplant follows later. The target estimand may then be any chosen contrast, e.g. the difference between the survival curves $S_1 (t) - S_0 (t) = P(T_1>t) - P(T_0>t)$. Often one simply focuses on the hazard ratio for treatment after adjusting for baseline covariates.    

In practice, we turn to our RRT incidence registry where the ability to receive PKT depends on the patient in need of treatment as well as the availability of a suitable organ \cite{Pradel2008Survey}. Virtually all patients receiving PKT at RRT, would  technically have the option to start treatment with dialysis. The other way around is less obvious. The average effect of PKT among the PKT-receivers (ATT) has therefore a more straightforward interpretation than the corresponding ATNT. Without knowing how to find a kidney for transplant at present, we may still aim to evaluate what would happen if the PKT treatment became available. This follows the philosophy on causal effects from Jan Vandenbroucke \cite{Vandenbroucke2016Causality} and Bradford Hill \cite{Hill1965Causality}. The ATE within the full cohort considered capable of receiving either treatment will be a weighted average of the ATT and ATNT. 

A different estimand of interest may indicate the survival time lost while being treated with dialysis when waiting for a later transplant, relative to $T_1.$  Instead of ignoring any delayed transplant as in the ITT, it considers observed time $T$ ($=T_0$) in the dialysis group as a sum of two observable variables: $T = T_w + T_r$, where $T_w$ is the survival time spent without initial transplant and $T_r$, the residual survival time following the delayed transplant (if any).

The estimand may then focus specifically on the amount of time spent without transplant and estimates its effect. This would resemble the \textbf{as-treated analysis} described by Danaei et al.\cite{Danaei2013StatinsEmulation} by focusing on the `total duration of treatment'. In our case, we model the potential survival under PKT, $T_1$ as a function of $T_w$ and $T_r$ as:  $T_1 \stackrel{d}{=} T_w \exp(-\psi) + T_r$. Every day on initial dialysis then counts the potential $\exp(-\psi)$ days on PKT. [e.g. when $exp(-\psi) = 2$ the time spent on dialysis could have been doubled by starting treatment with PKT]. This model lets the residual survival time to be unaffected by an initial trajectory with or without transplant.  

An alternative as-treated model, transforms both parts of the sum to reflect an additional impact of the delayed transplant, e.g. $T_1 \stackrel{d}{=} T_w \exp(-\psi_w) + T_r \exp(-\psi_r),$ where generally, the $\psi-$parameters could depend on other baseline factors, and the timing of the transplant. 

For completeness, we point to the effect of choosing to transplant at a given delay time $t_0$ post RRT, versus further delaying the transplant or even staying on dialysis throughout. To answer questions on the effect of transplant timing one needs, however,  measures on time-varying confounders of the timing of transplant in the dataset. Without such data, we set out to estimate the first two estimands described here, relying on the assumption of no unmeasured {\it baseline} confounders as explained in the following two sections. 

\section{Analyzing potential survival under each treatment}\label{Potential_survival}

In the real world, the comparison between $T_1$ and $T_0$ relies on estimating the survival of two groups of different patients with similar baseline covariates but who experienced different treatments. The NUBC given baseline covariates $\pmb{Z}$, assumes $ \{T_0, T_1\} \indep PKT | \pmb{Z}$: the vector of potential survival times is independent of observed PKT once baseline covariate values have been fixed. A causal effect can then be represented by contrasting $S_1(t;\pmb{Z}) = P(T_1 > t | \pmb{Z}) = P (T>t | PKT=1, \pmb{Z})$  with $S_0(t;\pmb{Z}) = P(T_0 > t | \pmb{Z}) = P (T>t | PKT=0, \pmb{Z})$. Under non-informative censoring, $T \indep  C| \{ \pmb{Z}, PKT \}$ with $C$ the censoring time, the conditional survival functions $P (T>t | PKT=p, \pmb{Z})$ can be estimated in the usual way, typically through proportional or additive hazards models. We will use the former here.

\subsection{Assessing positivity}\label{Method_Positivity}
For (sub)populations that are open to starting either treatment a meaningful population treatment effect can be estimated. To satisfy this, so-called, positivity assumption their propensity score distributions must overlap \cite{Petersen2012Positivity}. Subgroups with treatment propensity (close to) zero or one obviously violate the positivity assumption as they  represent groups with little chance of receiving one of the treatments. To check for this, a propensity score (PS) model was built using logistic regression for the probability of receiving PKT from baseline covariates age, sex, region, primary kidney disease and calendar year of RRT onset. Interactions between age and sex with primary kidney disease and calendar year of RRT onset were included. Notwithstanding generally good overlap of PS in both groups (Figure \ref{fig:PS}), we found that patients with cancer or older than 75 appeared to rarely receive PKT yielding a PS close to 0. We excluded them from the target population thus adjusting the scope of this analysis.  

\subsection{Survival summary of observed initial treatment}\label{KM}
We estimated survival from the date of RRT onset onwards censoring patients still alive by 31 December 2017, as confirmed by the cause of death registry. KM curves per treatment group are seen to present robust survival chances in selected observed treatment groups, provided non-informative censoring holds. This condition fails when conditioning on covariates $\pmb{Z}$ is required to render censoring time $C$ independent of the survival time: $C \indep T | \pmb{Z}$. This can easily happen in long term disease registers, if cohorts entering later differ in baseline prognostic factors and/or enjoy a better survival time (conditional on these baseline factors). In the early years, transplants were offered only to highly selected groups eg younger and healthier patients. Over time, this treatment option was extended to a broader group of patients. This structure introduces informative censoring for the unadjusted KM curves as later cohorts are censored earlier and have shorter survival. 

Aware of this problem, we start by presenting ``the usual'' KM curves for two observed groups: PKT and dialysis first and see a much higher curve for $PKT=1$ patients, possibly explained in part by baseline confounding and informative censoring. Yet a third source of bias enters when we consider the subset of PKT=0 patient who received delayed transplantation before study end. Figure \ref{fig:KM} shows the overestimation of survival in the dialysis arm when studying this selective subgroup and it also suggests the extent of the immortal time bias explained in Section \ref{Case-study}

Further insight into the dialysis first group is supported by showing the cumulative incidence of transplant and of death without transplant. Since the vast majority of patients experienced either competing event in this arm, any remaining informative censoring becomes negligible here. 

\subsection{Adjusting for confounders} \label{Confounder}

Not all the envisaged confounders (age, sex, region, primary kidney disease, calendar year of RRT onset, diabetes, hypertension, ischemic heart disease, cerebrovascular disease and peripheral artery disease) were always measured. Comorbidites were available only from 1998 onwards. Removing patients who entered before 1998 would result in losing 36\% of events, a substantial information loss.  Instead, we imputed the missing values for the earlier cohort assuming missingness at random, effectively extrapolating their conditional distribution from 1998 (allowing for a trend in calendar time) towards the early years. Similar patterns are found in long term chronic disease registers, that introduce additional covariates when the registers are already established. We decided not to impute kidney function because it was introduced later (only in 2008) and the reported measure was not standardised i.e. it is not a mandatory variable, centres can provide different measurement to report and different follow-up points. Instead, we assessed the impact of kidney function as part of the Sensitivity Analyses described in Section \ref{Sensitivity}.

Following Clark and Altman \cite{Clark2003MissingData}, we included the mortality indicator and log(survival time) in the imputation model as covariates, besides age, sex, region, primary kidney disease, calendar year of RRT onset and PKT. For computational efficiency, we first created 10 imputation datasets using the R package mice \cite{Van2015MICE} and then we bootstrapped each imputed dataset to construct 95\% CI using the R package boot \cite{Canty2020Boot} following previous recommendations \cite{Schomaker2018Bootstrap}.

To adjust for baseline confounders in survival analysis, one has in principle 3 options: regression adjustment, inverse probability weighting and/or a combination in a double robust method. We have opted for the regression adjustment because it allows us to automatically adjust for covariates known or suspected to affect censoring. For instance, by adjusting for calendar time of entry into the register we remove some informative censoring from the analysis. On the contraty, inverse probability weighting would balance covariates between the treatment groups, but observed hazards would still be subject to censoring that is influenced by baseline covariates or even time-varying covariates that are not available in the SRR. To correct for this we would need additional time-varying inverse weighting for censoring. Thus, our choice here aims at simplicity and robustness for the setting. Below we describe our modelling approach which is then compared with the IPW alternative to illustrate these considerations. 

We built Cox models for mortality separately in the PKT group and dialysis first group in each imputed dataset. The separate models give more flexibility, allowing for different baseline hazards and covariate effects for each treatment in a setting where there is potential for a different evolution over time. To avoid smoothing bias, we use a common set of confounders that are adjusted for in both models. As shown by the sensitivty analyses performed, the impact of the confounder adjustment on individual survival diminishes, once the curves are averaged over the population of interest.  

We then derived covariate-specific potential survival curves under each possible treatment ($\hat{S_1}[t|\pmb{Z}_i]$ and $\hat{S_0}[t|\pmb{Z}_i]$.  The average of these two curves over the whole study population was contrasted next to estimate the average treatment effect as $\hat{S_1}(t)-\hat{S_0}(t)$, with $\hat{S_1}(t) = \frac{1}{n} \sum_i \hat{S_1}[t|\pmb{Z}_i]$ and $\hat{S_0}(t) = \frac{1}{n} \sum_i \hat{S_0}[t|\pmb{Z}_i]$.

We similarly averaged over the covariate distribution observed in the PKT (and dialysis first) group to estimate the average treatment effect among the treated (and non-treated). If the model has adjusted for a sufficient set of baseline confounders, these results can be interpreted as causal effects under the potential outcomes framework for the targeted populations. We further assess residual confounding with the sensitivity analysis described below. Without relying on the no unmeasured confounding assumption, we are still contrasting well defined standardized survival curves. As secondary analysis, we repeated the analyses avoiding imputation by excluding comorbidities from the set of confounders.

To compare different adjusting approaches, we use the package ipw \cite{Van2011IPW} to build inverse probability of treatment survival curves, using the same baseline confounders as described above. For each of the 10 imputed datasets, we compute the weight for each patient and then averaged over the 10 sets to get the individual weight that was finally used in the curves.  

\subsection{Sensitivity analyses}\label{Sensitivity}

Our approach naturally involves three types of `untestable' assumptions, namely:
non-informative censoring, no unmeasured confounding and non-informative missingness for any other variables involved in the analysis. We consider the plausibility of each of these assumptions in turn and perform sensitivity analyses when questions arise, as described below.

Non-informative censoring, $T \indep C |\pmb{Z},PKT$ is required for any survival analysis (causal or not) and defined in function of the (baseline) covariates conditioned upon. We argue that the cohorts who entered the registry in more recent calendar years may have better survival because  care generally improves over the years. As highlighted before, the risk profiles of patients entering the cohort also change over time. Hence any analysis which fails to adjust for calendar time (or a sufficient proxy) may suffer from bias due to informative censoring. To illustrate the impact here, we compared the `nonparametric' KM curve, with a standardized curve (averaging over covariate adjusted survival). We anticipate that the latter curve will demonstrate better survival as it is less dominated by the early cohort entries which have the longer follow-up time and add more events to our study. We note that an IPW weighted KM curve, which may involve predictors of survival time, does not remedy for this as we will explain in section \ref{Results} and could make things even worse (as we found out). The IPW version may benefit from using time-varying weights but, as already stated, time-varying confounders are not available in the registry. 

Regarding the no unmeasured confounding assumption, we are limited to what is registered and since when. Prognostic factors identified in previous studies, as comorbidites and kidney function are not available for the full cohort but only introduced in 1998 and 2008, respectively \cite{Milton2008Australia,Haller2017Austria}. We have chosen to check the impact on the targeted marginalized survival curves of adding these or not - using the data in the respective calendar windows where they are available. This resulted in involving comorbidities in our analyses, after imputing them for the 1991-1998 period, but ignoring the GFR for our full cohort analyses.

To consider unmeasured confounders, we looked at the strength that one additional unmeasured confounder would need to have in order to qualitatively change the current conclusion. First, we repeated the estimation procedure on a subsample for which we have access to registered baseline kidney function. These new models included the covariates listed before plus kidney function as main effect and as interaction with age and sex. We then compared the survival estimated effect, with the effect derived from models that drop age as a covariate, given that age is a well-known prognostic factor.  

Finally, our analyses rely on missingness at random for the multiple imputation approach to be valid. An alternative approach is to limit the assessment to the full cases. Thus, we considered the initial set of potential confounders on the complete cases dataset: patients who started RRT in 1998 or later whose comorbidities are registered and derive their standardised survival curves. We then examined how the standardised estimates for this subgroup change when comorbidities are dropped from the covariate list. 

\section{Time lost while on initial dialysis} \label{Methods_AFT}

\begin{figure}[hbt!]
    \centering
    \includegraphics[width=0.8\textwidth]{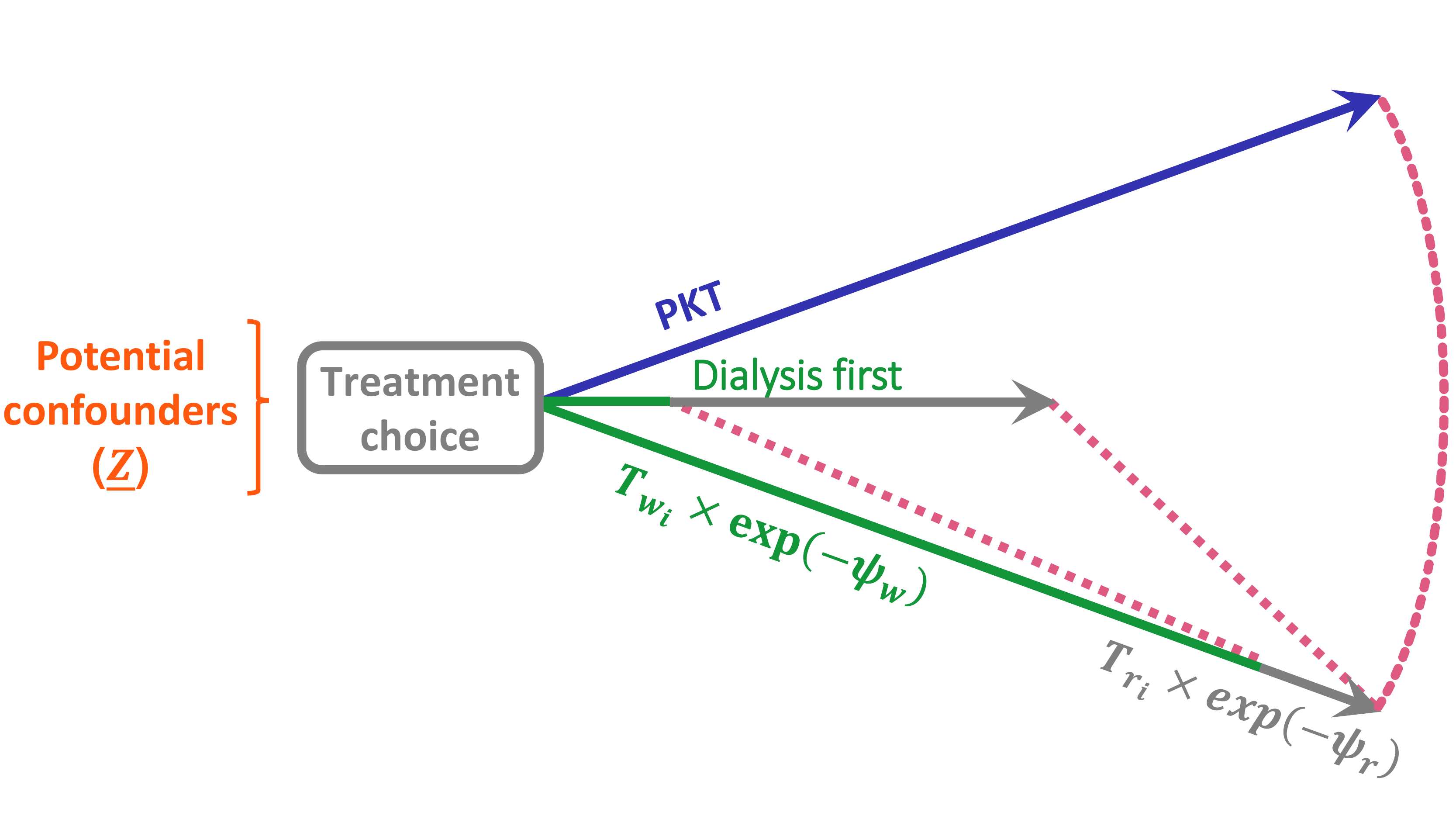}
    \caption{Structural accelerated failure model that relates survival time on dialysis to what it might have been following PKT}
    \label{fig:AFT}
\end{figure}

Patients starting on dialysis continue with it for different lengths of time before possibly switching to transplant. To estimate the impact on survival  of the time spent on initial dialysis we invoke the structural accelerated failure time (AFT) model illustrated in Figure \ref{fig:AFT}.

A one parameter AFT model transforms the initial time spent without transplant to what it might have been under PKT, and leaves the residual time on transplant unchanged. Specifically, the observed time without transplant  $T_{w}$ is multiplied by a factor $\exp(-\psi)$ and then added to the observed time without transplant, $T_{r}$ to arrive at $T_{1} (\psi)$ 
the potential survival time from RRT onset to death under PKT:

$$T_1 (\bpsi) \stackrel{d}{=} T_w \exp (-\psi) + T_r $$

The model expresses that a day on initial dialysis would have amounted to  $\exp(-\psi)$ days had the patient received PKT instead. A longer survival time under the PKT scenario corresponds to a negative value of $\psi$.

More generally 
$$T_1 (\bpsi) \stackrel{d}{=}   \in_0^T   \exp(\bpsi) D_i(u) du, $$ where $D_i (u) $ indicates whether individual $i$ is still on initial dialysis at time $u$ or not \cite{Latimer2020G-estimation}.

A still simple but more flexible model allows for an altered “transplant effect” when to follows a period on dialysis rather than PKT:
$$T_1 (\bpsi) \stackrel{d}{=} T_w \exp (-\psi_w) + T_r \exp (-\psi_r)  $$
where the factor $\exp(-\psi_r)$ now backtransforms residual time after transplant, $T_r$, in addition to  the factor for “immediate transplant effect” of PKT, $\exp(-\psi_w)$, that multiplies the time without dialysis $T_w$ as in Figure \ref{fig:AFT}.

Parameters in this model can be obtained through G-estimation relying on the NUBC assumption \cite{Hernan2005AFT}. Upon  transforming the observed initial time on dialysis $T$ to what it would have been under PKT $T_1(\bpsi)$ we obtain potential survival times. For the true parameter $\bpsi^*$ the transformed potential time $T_0 \stackrel{d}{=} T_1(\bpsi^*)$ no longer depends on the observed treatment PKT, once the necessary baseline confounders have been accounted for. We fitted a Cox PH model, regressing $T (\bpsi) $ on $\pmb{Z}$ with an additional  effect of PKT  on the log hazard scale.  The value of $\bpsi$ 
for which the involved 
coefficient $\hat{\bbeta}_{PKT}
(\bpsi) $ is zero is our point estimate.
For the 2 parameter case, the estimation model involved an interaction between PKT and sex. A Wald test for $\hat{\bbeta}_{PKT}
(\bpsi)=0$ was used to identify the  value of the parameter vector $\bpsi$ with 95\% confidence intervals. 

As the backtransformed censoring time depends on the covariates, we repeated this estimation process using artificial censoring to avoid informative censoring \cite{White1998AFTcensoring}. Thus, the backtransformed time under each potential $\psi$ was administratively censored after 5, 10, 15 years and 20 years. However, the earlier the administrative censoring occurs, the more information we lose as fewer events remain. Identifying a new $\bpsi^*$ under each of these censoring cutoffs that disregard different proportions of events, provides evidence on the robustness of the estimate for the original model.

\section{Results}\label{Results}

\begin{table}[hbt!]
    \centering
    \begin{tabular}{|l|r|r|}
        \hline
        Study population & PKT & Dialysis first\\
        \hline
        Number of adult patients from SRR 1991-2017 & 1,214 (100.0) & 28,312 (100.0)\\
        \hline
        Number of patients older than 75 years & 4 (0.3) & 7,108 (25.1)\\
        \hline
        Number of patients from foreign or unknown region & 18 (1.5) & 170 (0.6)\\
        \hline
        Number of patients who receive RRT abroad & 57 (4.7) & 79 (0.3)\\
        \hline
        Number of patients who died or got censored on same day of RRT onset & 1 (0.1) & 20 (0.1)\\
        \hline
        Number of patients with a history of cancer or unknown cancer status & 37 (3.0) & 2,501 (8.8)\\
        \hline
        Total sample & 1,097 (90.4) & 18,434 (65.1)\\
        \hline
        \end{tabular}
    \caption{Study population selection and number of individuals related to exclusion criteria}
    \label{tab:Exclusion}
\end{table}

\begin{figure}[hbt!]
    \centering
    \includegraphics[width=12cm]{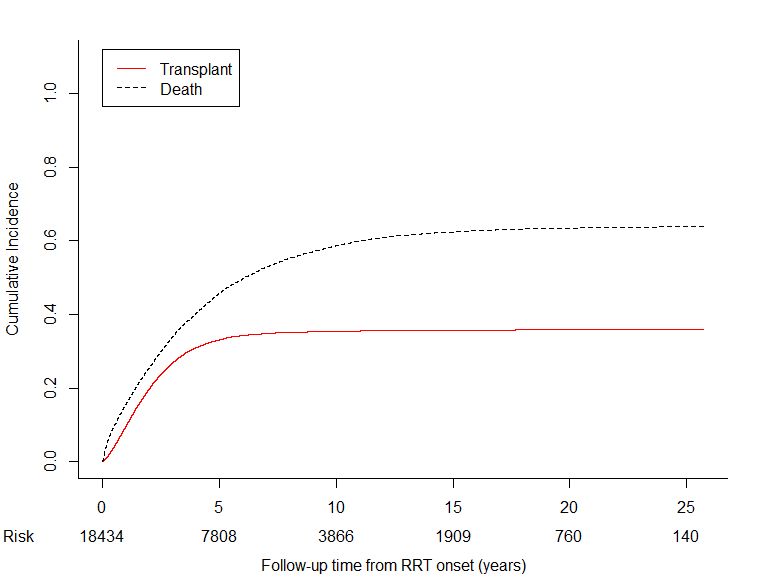}
    \caption{Cumulative incidence of transplantation and death without transplanation in the dialysis first group}
    \label{fig:Cumulative_incidence}
\end{figure}

By December 2017, the SRR included 29,526 adult patients of whom 1,214 started with PKT. After excluding patients older than 75 years, non-Swedish residents, those who received RRT abroad, and those who died on the day of RRT onset or had a history of cancer, the study population included 1,097 PKT and 18,434 dialysis first patients (Table \ref{tab:Exclusion}). The median time on initial dialysis, prior to transplant, death or censoring, was 2 years (Figure \ref{fig:Cumulative_incidence}). There were more deaths observed in the dialysis group compared to the PKT group. 
Table \ref{tab:Survival} summarises the survival outcomes.

\begin{table}[hbt!]
    \centering
    \resizebox{0.8\textwidth}{!}{\begin{tabular}{|l|r|r|c|c|c|}
        \hline
          & & & & Median & \\
          (Sub)population & Patients, & Deaths & \% deaths  & person-years & Hazard \\
          & n (\%) & n (\%)& per row & at risk & rate\\
        \hline
        RRT cohort & 19,531 (100) & 12,073 (100) & 61.8 & 4.1 & 0.10\\
        PKT group & 1,097 (5.6) & 196 (1.6) & 17.9 & 7.6 & 0.02\\
        Dialysis first group & 18,434 (94.4) & 11,877 (98.4) & 64.4 & 3.9 & 0.11\\
        \hline
    \end{tabular}}
    \caption{Survival summary}
    \label{tab:Survival}
\end{table}

Table \ref{tab:Covariate} describes covariate distribution in the overall RRT cohort and the PKT and dialysis first groups. PKT patients were younger and had less comorbidities than dialysis first patients. The distribution of primary kidney disease also differed between the groups. Hence, the need to adjust for confounding. 

\begin{figure}[hbt!]
    \centering
    \includegraphics[width=12cm]{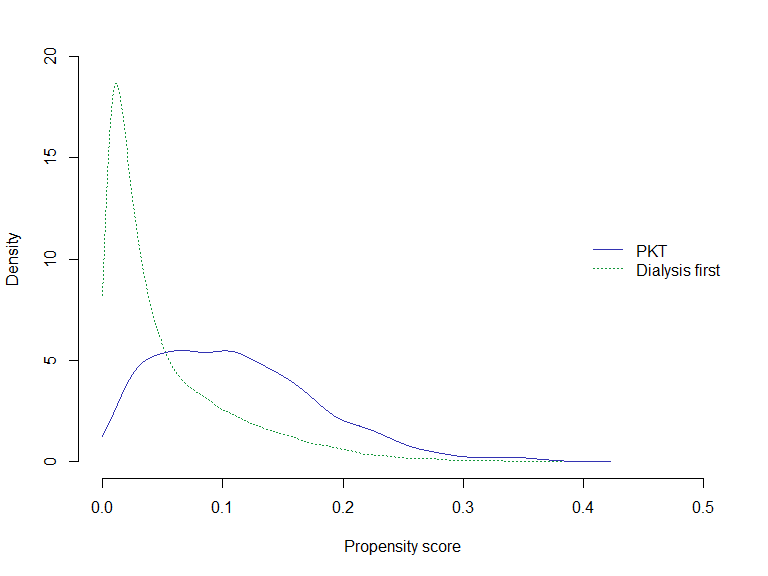}
    \caption{PKT propensity score density plot in the two observed treatment groups in the final study population.}
    \label{fig:PS}
\end{figure}

To assess the positivity assumption, we built a PS score model for PKT. Figure \ref{fig:PS} shows overlap in the PS for PKT between patients who effectively receive PKT and those who started on dialysis.

\begin{sidewaystable}
    \centering
    \resizebox{\textwidth}{!}{\begin{tabular}{|l|r|r|r|r|}
        \hline
          Covariate & RRT & 1) PKT & 2) Dialysis first  & Difference between \\
          & (n=19,531) & 1) (n=1,097) & 2) (n=18,434) & 1 and 2 (95\% CI)\\
        \hline
        Age, median (IQR) & 60 (20) & 47 (22) & 61 (19) & -12  ( -13 , -12 )\\
        Sex (female), n (\%) & 6,867 (35.2) & 411 (37.5) & 6,456 (35.0) & 2.44 (-0.6, 5.4)\\
        Region (Stockholm, reference), n(\%) & 3,649 (18.7) & 193 (17.6) & 3,456 (18.7) & -1.15 (-3.5, 1.2)\\
        Region (Uppsala/Orebro), n (\%) & 4,504 (23.1) & 249 (22.7) & 4,255 (23.1) & -0.38 (-3.0, 2.2)\\
        Region (Northern), n (\%) & 2,017 (10.3) & 95 (8.7) & 1,922 (10.4) & -1.77 (-3.5, 0.0)\\
        Region (Southern), n (\%) & 3,466 (17.7) & 179 (16.3) & 3,287 (17.8) & -1.51 (-3.8, 0.8)\\
        Region (Southeastern), n (\%) & 2,376 (12.2) & 127 (11.6) & 2,249 (12.2) & -0.62 (-2.6, 1.4)\\
        Region (Western), n (\%) & 3,519 (18.0) & 254 (23.2) & 3,265 (17.7) & 5.44 (2.8, 8.0)\\
        Kidney disease (Diabetic nephropathy, reference), n (\%) & 5,656 (29.0) & 183 (16.7) & 5,473 (29.7) & -13.01 (-15.4, -10.7)\\
        Kidney disease (Glomerulonephritis), n (\%) & 3,508 (18.0) & 328 (29.9) & 3,180 (17.3) & 12.65 (9.8, 15.5)\\
        Kidney disease (Uremia of unknown cause), n (\%) & 2,063 (10.6) & 116 (10.6) & 1,947 (10.6) & 0.01 (-1.9, 1.9)\\
        Kidney disease (Polycystic kidney disease), n (\%) & 1,538 (7.9) & 165 (15.0) & 1,373 (7.4) & 7.59 (5.4, 9.8)\\
        Kidney disease (Pyelonephritis), n (\%) & 640 (3.3) & 41 (3.7) & 599 (3.2) & 0.49 (-0.7, 1.7)\\
        Kidney disease (Other), n (\%) & 6,126 (31.4) & 264 (24.1) & 5,862 (31.8) & -7.73 (-10.4, -5.1)\\
        
        Hypertension*, n (\%) & 15,520 (79.5) & 832 (75.8) & 14,688 (79.7) & -3.8 (-4.7, -3.0)\\
        Diabetes*, n (\%) & 7,405 (37.9) & 202 (18.4) & 7,203 (39.1) & -20.7 (-21.5, -19.9)\\
        Ischemic heart disease*, n (\%) & 5,196 (26.6) & 49 (4.4) & 5,147 (27.9) & -23.5 (-23.9, -23.0)\\
        Peripheral artery disease*, n (\%) & 2,582 (13.2) & 33 (3.0) & 2,550 (13.8) & -10.8 (-11.2, -10.5)\\
        Cerebrovascular disease*, n (\%) & 2,072 (10.6) & 22 (2.0) & 2,050 (11.1) & -9.1 (-9.4, -8.8)\\
        \hline
        Outcome: Deaths, n (\%) & 12,073 (61.8) & 196 (17.9) & 11,877 (64.4) & -46.5 (-49, -44.1)\\
        \hline
        
    \end{tabular}}
    \caption{Covariate distributions over the observed treatment groups\\
    \scriptsize{*Imputed covariates. The mean over the 10 imputed datasets is presented}}
    \label{tab:Covariate}
\end{sidewaystable}

\begin{figure}[hbt!]
    \centering
    \includegraphics[width=\textwidth]{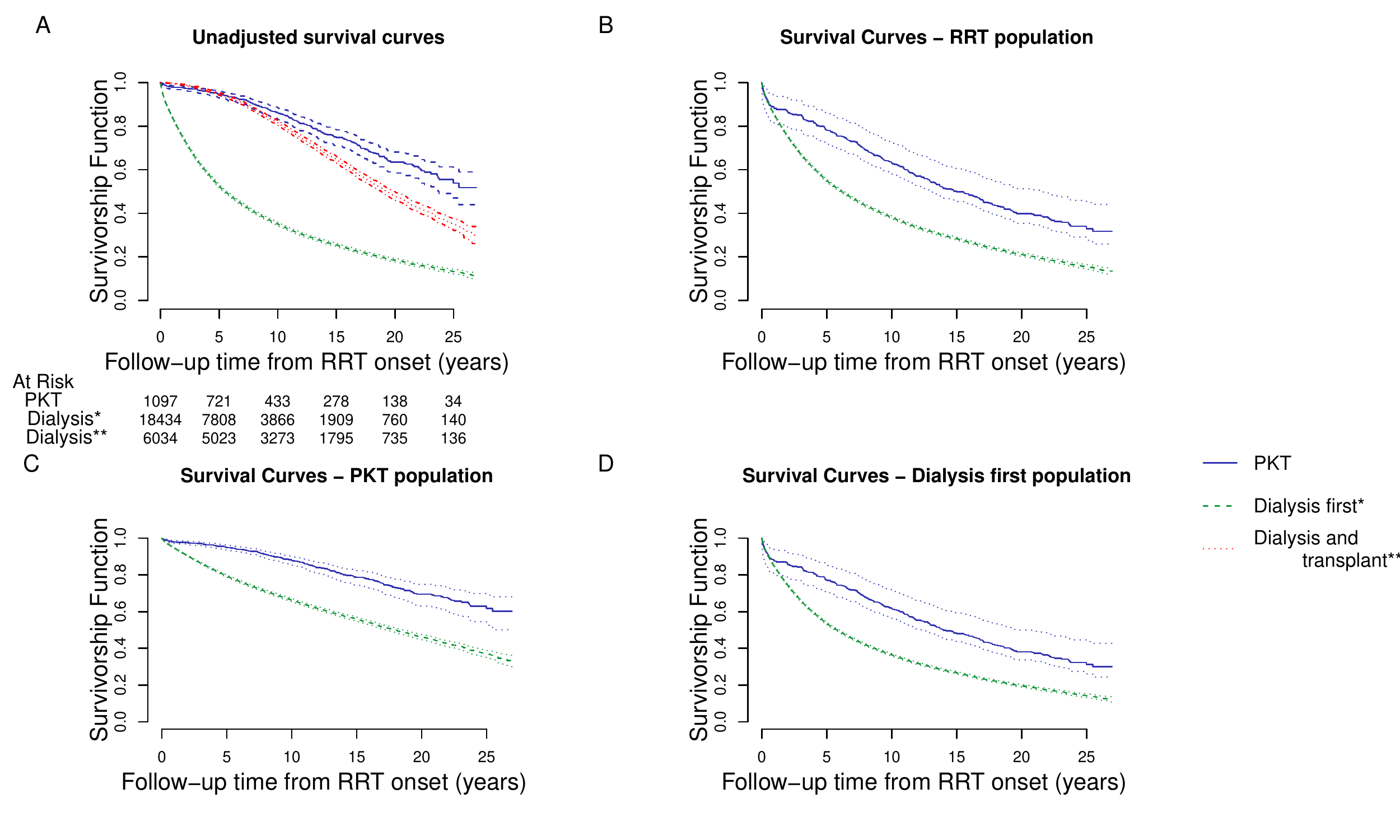}
    \caption{Unadjusted and adjusted survival curves from RRT onset. A. Unadjusted Kaplan Meier curves of observed mortality are shown for the PKT subpopulation (blue), the dialysis first subpopulation (green) and the subset of the dialysis first who received a transplant (red). These Kaplan-Meier curves are typically seen as robust estimates of the underlying population survival chances. Here, however, we discovered that calendar time of study entry predicts mortality and hence the unadjusted curves suffer from informative censoring. B. Model based survival curves under each potential treatment given the full RRT population covariates. C. Model based survival curves under each potential treatment given the PKT subpopulation covariates D. Model based survival curves under each potential treatment given the dialysis first subpopulation covariates.}
    \label{fig:KM}
\end{figure}

In Figure \ref{fig:KM}A, unadjusted KM curves from RRT onwards for the PKT and dialysis first group show better survival in the observed PKT group over dialysis first (log-rank test p<0.001). Since calendar time of study entry predicts mortality as described in Section \ref{KM}, these unadjusted curves suffer from informative censoring however. In addition, the curve for the subset of dialysis first patients who were seen to receive a later transplant suffers from immortal time bias. It is dramatically shifted upwards leaving no apparent difference with the PKT-curve for the first 5 years. Estimating survival on this selective subset clearly leads to overestimation of survival in the dialysis arm.  

To account for baseline confounders, standardised survival curves were built. Figure \ref{fig:KM} shows patient survival after RRT onset, derived from the corresponding models as described in Section \ref{Confounder}: once averaged over the covariates of the whole RRT cohort and then over the PKT and the dialysis first groups separately. The PKT survival advantage over the dialysis first strategy, appears only partially explained by differences in baseline variables, e.g. being younger and healthier. 

Table \ref{tab:Average_survival} summarizes patient survival in each (sub)population at different time points. The population specific risk differences reveal dramatical treatment impact over the PKT subpopulation, the dialysis first subpopulation and the full RRT population. For the RRT and the dialysis first populations, the biggest estimated difference in survival occurs between 5 and 10 years after RRT onset. For the PKT group this difference continues to increase over follow-up time. PKT patients form a highly selective group that enjoys better survival under either treatment. 

\begin{table}[hbt!]
    \centering
    \resizebox{0.9\textwidth}{!}{\begin{tabular}{cccc}
    \hline
    Year(s) after & Survival under PKT & Survival under dialysis first & Difference in survival \\
    RRT onset & (95\% CI) & (95\% CI) & (95\% CI)\\
    \hline
    RRT population \\
    \hline
    1 & 0.88 (0.81, 0.94) & 0.85 (0.84, 0.85) & 0.03 (-0.04, 0.09)\\
    
    5 & 0.78 (0.72, 0.86) & 0.55 (0.54, 0.56) & 0.23 (0.17, 0.31)\\
    
    10 & 0.63 (0.58, 0.73) & 0.38 (0.38, 0.39) & 0.25 (0.20, 0.35)\\
    
    15 & 0.50 (0.45, 0.60) & 0.28 (0.28, 0.29) & 0.22 (0.17, 0.32)\\
    
    20 & 0.40 (0.35, 0.51) & 0.21 (0.20, 0.22) & 0.19 (0.14, 0.30)\\
    
    25 & 0.33 (0.28, 0.45) & 0.15 (0.14, 0.16) & 0.18 (0.12, 0.30)\\
    \hline
    PKT subpopulation \\ 
    \hline
    1 & 0.98 (0.97, 0.99) & 0.94 (0.94, 0.95) & 0.03 (0.03, 0.04)\\
    
    5 & 0.95 (0.94, 0.96) & 0.79 (0.79, 0.80) & 0.16 (0.14, 0.17)\\
    
    10 & 0.88 (0.86, 0.90) & 0.66 (0.65, 0.67) & 0.22 (0.19, 0.24)\\
    
    15 & 0.79 (0.74, 0.82) & 0.56 (0.55, 0.57) & 0.23 (0.18, 0.27)\\
    
    20 & 0.70 (0.63, 0.75) & 0.46 (0.45, 0.48) & 0.23 (0.17, 0.29)\\
    
    25 & 0.62 (0.53, 0.69) & 0.37 (0.35, 0.39) & 0.25 (0.16, 0.33)\\
    \hline
    Dialysis subpopulation \\
    \hline
    1 & 0.88 (0.80, 0.94) & 0.84 (0.84, 0.85) & 0.03 (-0.04, 0.09)\\
    
    5 & 0.77 (0.71, 0.85) & 0.54 (0.53, 0.54) & 0.24 (0.17, 0.32)\\
    
    10 & 0.62 (0.57, 0.72) & 0.36 (0.36, 0.37) & 0.25 (0.20, 0.35)\\
    
    15 & 0.48 (0.44, 0.59) & 0.27 (0.26, 0.27) & 0.21 (0.17, 0.32)\\
    
    20 & 0.38 (0.34, 0.50) & 0.20 (0.19, 0.20) & 0.19 (0.14, 0.30)\\
    
    25 & 0.31 (0.26, 0.44) & 0.14 (0.13, 0.15) & 0.17 (0.12, 0.30)\\
    \hline
    \end{tabular}}
    \caption{Survival probabilities for the different (sub)groups of interest under the two potential treatments: pre-emptive kidney transplantation (PKT) and dialysis first derived from the model with imputed covariates}
    \label{tab:Average_survival}
\end{table}

Table \ref{tab:Average_survival_simple} summarises the patient survival in each (sub)population at different time points when estimated from the models built as secondary analysis i.e. they did not include comorbidites as confounders. Compared to the estimates from the full original model, the estimated survival under dialysis for the different (sub)populations is essentially the same. However, the estimated potential survival under PKT for the dialysis and RRT groups is slightly higher from these models compared to the estimates derived from the models including comorbidities.

\begin{table}[hbt!]
    \centering
    \resizebox{0.9\textwidth}{!}{\begin{tabular}{cccc}
    \hline
    Year(s) after & Survival under PKT & Survival under dialysis first & Difference in survival \\
    RRT onset & (95\% CI) & (95\% CI) & (95\% CI)\\
    \hline
    RRT population \\
    \hline
    1 & 0.92 (0.87, 0.95) &	0.85 (0.84, 0.85) &	0.07 (0.02, 0.10)\\
    
    5 & 0.82 (0.78, 0.87) &	0.55 (0.54, 0.56) &	0.27 (0.23, 0.32))\\
    
    10 & 0.67 (0.62, 0.72)	& 0.38 (0.38, 0.39)	& 0.28 (0.23, 0.34)\\
    
    15 & 0.52 (0.46, 0.58) & 0.28 (0.28, 0.29) &	0.23 (0.18, 0.30)\\
    
    20 & 0.41 (0.35, 0.48) & 0.21 (0.20, 0.22) & 0.20 (0.14, 0.27)\\
    
    25 & 0.33 (0.27, 0.41) & 0.16 (0.14, 0.17) & 0.18 (0.11, 0.26)\\
    \hline
    PKT subpopulation \\ 
    \hline
    1 & 0.98 (0.97, 0.99) & 0.94 (0.94, 0.94) & 0.04 (0.03, 0.05)\\
    
    5 & 0.95 (0.94, 0.96) & 0.78 (0.78, 0.79) & 0.17 (0.15, 0.18)\\
    
    10 & 0.88 (0.86, 0.90) & 0.65 (0.64, 0.66) &    0.23 (0.20, 0.25)\\
    
    15 & 0.78 (0.74, 0.82) & 0.55 (0.54, 0.56) & 0.23 (0.19, 0.27)\\
    
    20 & 0.69 (0.63, 0.74) & 0.46 (0.44, 0.47) & 0.23 (0.17, 0.29)\\
    
    25 & 0.61 (0.53, 0.68) & 0.37 (0.35, 0.39) & 0.25 (0.16, 0.32)\\
    \hline
    Dialysis subpopulation \\
    \hline
    1 & 0.91 (0.87, 0.95) & 0.84 (0.84, 0.85) & 0.07 (0.02, 0.10)\\
    
    5 & 0.82 (0.77, 0.86) & 0.54 (0.53, 0.54) & 0.28 (0.23, 0.33)\\
    
    10 & 0.65 (0.60, 0.71) & 0.37 (0.36, 0.37) & 0.29 (0.24, 0.35)\\
    
    15 & 0.50 (0.45, 0.57) & 0.27 (0.26, 0.28) & 0.23 (0.18, 0.30)\\
    
    20 & 0.39 (0.34, 0.47) & 0.20 (0.19, 0.21) & 0.19 (0.14, 0.27)\\
    
    25 & 0.32 (0.25, 0.40) & 0.14 (0.13, 0.15) & 0.17 (0.11, 0.26)\\
    \hline
    \end{tabular}}
    \caption{Survival probabilities for the different (sub)groups of interest under the two potential treatments: pre-emptive kidney transplantation (PKT) and dialysis first derived from the model without comorbities}
    \label{tab:Average_survival_simple}
\end{table}

\begin{figure}[hbt!]
    \centering
    \includegraphics[width=12cm]{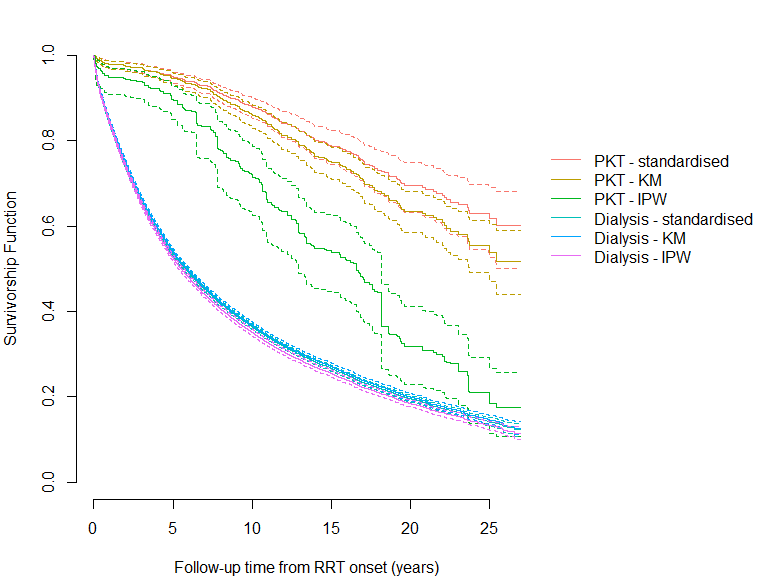}
    \caption{Comparison of unadjusted and adjusted survival curves through standardisation and IPW}
    \label{fig:IPW}
\end{figure}

Figure \ref{fig:IPW} shows unadjusted and adjusted survival curves for the PKT and dialysis population using standardisation and IPW. For the dialysis group, the curves overlap. For the PKT group, standardisation yields better survival than the unadjusted KM curves, which in turns exceeds the estimated survival with IPW.   

Figure A and B, on Appendix, show the standardised survival curves when omitting age from the original models for both the PKT and dialysis first groups as a sensitivity analyses. The curves are quite different particularly when estimating the counterfactual outcome i.e. the treatment that was not observed. However when this was repeated for the subset of patients that had additional covariates, first for those with comorbidities reported (C and D) and then for those with baseline kidney function (E and F) there was quite an overlap with the estimated survival curves without these additional confounders suggesting little impact of omitting such variables (Appendix). 

A structural accelerated failure model with one parameter was built to account for time under dialysis. The corresponding acceleration parameter was $\exp (-\hat{\psi})$ = 4.8 (95\% CI 3.9, 5.8). Therefore assuming no unmeasured confounding, the survival time after PKT was almost 5-fold the survival time while on dialysis. In other words, on average every year of survival under initial dialysis is equivalent to surviving 4.8 years under PKT.

\begin{figure}[hbt!]
    \centering
    \includegraphics[width=12cm]{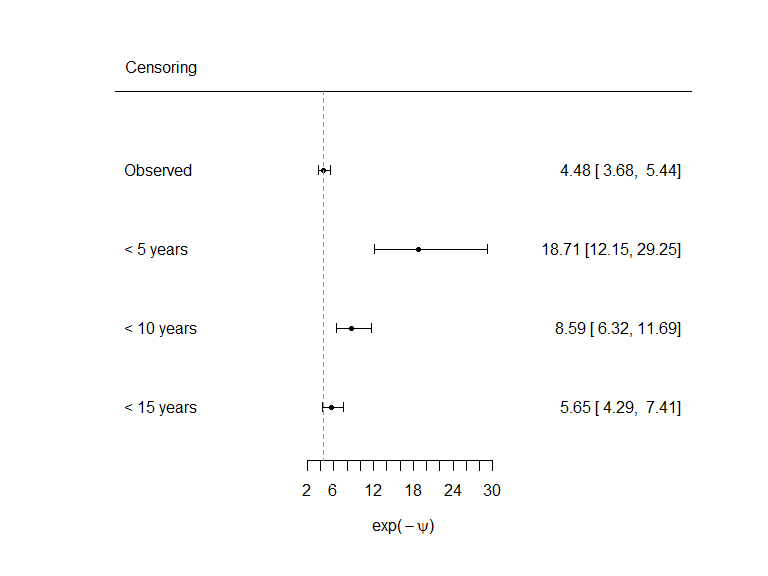}
    \caption{Comparison of different administrative censoring cutoffs for the estimated backtransformed survival time}
    \label{fig:AFT_censoring}
\end{figure}

Figure \ref{fig:AFT_censoring} shows the impact of different administrative censoring cutoffs for the backtransformed survival time and the corresponding number of events observed under each censoring scenario. As expected, the earlier the censoring occurs, the wider the CI given that fewer events are observed. More importantly, the estimated $\exp (-\hat{\psi})$ under the different administrative censoring scenarios consistently suggest a survival advantage of PKT, with the corresponding CI overlapping with that of the CI derived from the full observed data.

A second structural accelerated failure model with two parameters was built to account both for time under dialysis and beyond delayed transplantation. The corresponding acceleration parameters were $\exp (-\hat{\psi}_w)$ = 5.6 (95\% CI 5.1, 6.6) and $\exp (-\hat{\psi}_r)$ = 0.7 (95\% CI 0.5, 0.9). On average, one year survival under initial dialysis is equivalent to surviving 5.6 years under PKT and then a year of delayed transplant would be equivalent to living 0.7 years under PKT. 

\section{Using available software}\label{Software}
Existing software facilitates the application of both standardisation and IPW but may not fit one's set up. The R package 'stdReg' implements standardisation using Cox models \cite{Sjolander2016Standard}. The population that it standardises the estimates to, should be a subset of the one used to fit the regression model, which is not our case. Even if we were willing to fit a single model and include treatment as a covariate and `lose' the flexibility of having different covariate effects, the package does not allow for strata which means that we would also have to assume the same baseline hazard under each treatment. 

As already stated, we use the R package ipw to fit the IPW curves \cite{Van2011IPW}. However when deriving the weights, the model can have at most 20 covariates. This was exceeded with the number of covariates and interactions we adjusted for. Therefore, we evaluated the use of different subsets of the covariates included in the main model, keeping the main effects and including different interactions each time to complete the maximum of 20 covariates adjusted for. The different combinations yielded similar results as the ones already described (data not shown). We decided to present these curves as they were useful to illustrate that IPW curves may not be valid in this context due to informative censoring and lack of measured time-varying confounders. 

\section{Discussion}\label{Discussion}

This case study showed potential and limitations of exploiting a nationwide incident disease register with {\it baseline} covariates measured across centers to produce real world evidence on PKT and its effect on mortality of ESRD patients. It demonstrated how such registers more generally enable estimation of the total effect of a well-defined point exposure on survival time, provided a sufficient set of baseline confounders for the point exposure has been measured across the registry. Under additional semi-parametric modelling assumptions, we further estimated the effect of observed time spent off-initial-treatment allowing the `on switch’ to be affected by unmeasured time-varying covariates. We saw that careful consideration of the target estimand has major importance as results differed vastly over the relevant options. We finally presented estimation approaches that are well adapted to the setting and discussed their assumptions in context. 

Specifically, this case study has quantified to what extent PKT comes with higher standardized survival  over the period 1991-2017 for the Swedish AT(N)T (sub)populations considered. Assuming no unmeasured confounding at baseline, this represents a causal difference in survival chances. Prognostic factors differed greatly over the observed treatment groups (PKT versus dialysis at ESRD). With all available confounders in a relatively simple Cox model, standardized marginal survival revealed a large potential survival benefit in both  the observed PKT and Dialysis first populations. A more complex model could have been fitted, but we saw little impact of this for our outcome.  

In the PKT group (Figure \ref{fig:IPW}) a striking difference emerged between the adjusted survival curve obtained through Cox regression and through the corresponding IPW weighted Kaplan-Meier curve. This was explained (in part) by a calendar time trend in baseline risk as well as measured risk profiles, a common phenomenon in long term disease registries with complex treatments. Later cohort entry  (RRT onset) translates into  shorter administrative censoring time, while later cohorts show more older patients with additional comorbidities receiving PKT. IPW weighted KM curves work with covariate balance between treatment groups, but still rely on the hazard beyond censoring being well represented by that of uncensored patients. Since non-informative censoring does not hold, we must expect underestimated survival. The bias is greater even than with unweighted KM-curves as IPW upweights patients with older age who entered later into the registry and were therefore censored sooner. The IPW approach could work when a sufficient set of time-varying covariates were measured regularly to model the hazard of censoring and additional inverse time-varying probability of censoring were applied.  It would then of course involve a second model besides the propensity of treatment model. It is also extremely demanding and costly to measure regular time-varying covariates across the nation, and hence unrealistic in this case.

Not only risk profiles of  patients entering over time, but also treatment strategies are evolving. In our context surgical techniques were refined and newer and better immunosupression therapy was introduced. When it comes to predicting benefits for future patients, extrapolation should incorporate these trends. 

In the AFT analysis, we assessed the impact of immediate transplant vs delayed transplant on patient survival by estimating the time lost under initial dialysis. This provides further evidence that survival under PKT is better, regardless of the time under initial dialysis. Interestingly, for the AFT with 2-parameters, each parameter shows an estimated effect in opposite directions. But in any case, survival time gained beyond `delayed transplant’ does not outweigh the disadvantage posed by initial dialysis, where the order of magnitude of time lost under initial dialysis is almost 5-fold. In the absence of measured time-varying confounders, usually required for deriving as-treated effects, the approach here developed provides empirical evidence for decision making  applicable to other settings. 

Looking at the applied literature, we found it largely ignores many fundamental statistical lessons learned. This greatly hampers interpretability in our setting and a fortiori  transportability to new settings or the relevance of meta-analyses \cite{Vo2019Heterogeneity}. We hope the case study here developed will help support a change in statistical practice and inspire further research on challenges encountered. Below, we respond to some arguments often raised to justify suboptimal analysis and critically reflect on remaining challenges when targeting more explicit causal effect estimation as we did.  

Researchers (and editors) see no need to `refine’ the approach since the large positive outcome difference  for PKT leaves ample room for error before qualitative conclusions change. We found, however,  that restriction to the subset of dialysis starters selected upon delayed transplantation, virtually annuls the large survival difference for the full population of dialysis starters (Figure \ref{fig:KM}). Moreover, with higher risk profiles for kidney transplantation over calendar time, we must anticipate future study populations with smaller magnitudes of effect. 

Another argument against careful `causal’ analysis is its complexity which might deprive clinicians from a critical understanding of the opportunities and risks involved. We agree that several assumptions play a key role and must be discussed with clinicians. We found the tool of trial emulation, to enhances both insight in the data structure and interpretability of results for a broader scientific audience.  Understanding association is clearly  more simple but suffers when one jumps all to easily to causal conclusions. As \textbf{the critical assumption of NUCB} is fundamentally untestable, derived causal effect estimates should be complemented with an analysis of their sensitivity to various plausible violations. To this end, the helpful concept of an e-value deserves further development in the survival setting \cite{Vanderweele2017Evalue}. We may indeed have missed relevant confounders that account for part of the observed difference in survival between treatment groups as we discuss next. 

For our study, comorbidities and kidney function were only available from 1998 and 2008 onwards, respectively. Notwithstanding their significant effect on the patient specific hazards (Appendix \ref{App}), in these cohorts the population average survival remained essentially unchanged with or without adjustment for the newly available variables. 

We lacked socio-economic factors, while patients with higher socio-economic status could have better access to health care and timely treatment with higher chances of receiving PKT. If so, our estimated PKT advantage remains confounded by socio-economic factors and may be overestimated. Additional unmeasured confounders may include: unmeasured transient comorbidities such as current infections possibly delaying transplantation, patient preferences or availability of a live donor \cite{Pradel2008Survey,Ayanian1999Preferences}. The advantage of the Swedish system of registries with a unique patient identifier is that additional variables can be obtained in the future from further linkage. 

Focusing on the \textbf{ITT effect} of PKT versus dialysis first, we averaged over observed follow-up strategies as currently implemented. The PKT group then covers the `natural’ mix of cadaveric and living donor kidneys. To evaluate the impact on survival of donor type (and other treatment refinements)  one could treat `PKT from a living donor’ as the specific treatment of interest and study its benefit along the lines established in this paper \cite{Vanderweele2013MultipleVersions}. 

Potential survival under dialysis as estimated from the original models barely changed when using models ignoring covariates needing imputation. Potential survival under PKT however did change, particularly in the observed dialysis and full RRT group. By including comorbidities we accounted for worse baseline prognosis in the dialysis group and reduced the estimated benefit under PKT. Even so, PKT predicted better survival  in both the PKT and dialysis treated. With long term registers that gather periodic information across different centres, there is an unavoidable risk of missing data that may have influenced the estimated effect. It is good practice to enter new patient characteristics in an established registry when their role becomes apparent,  or when new diagnostic tools or treatment options are introduced. Considering the importance of incident disease registries as a resource for research providing real world evidence, we feel collaborative efforts should aim to define a minimum set of confounders to be reported. This will enhance transportability of results derived from single registry-based studies and improve evidence synthesis in meta-analytic approaches. 

For our case-study, we included all available confounders considered clinically relevant. Notwithstanding significant contributions to the Cox model, the impact of some covariates on the standardized curves  was limited, as shown by the sensitivity analyses. In settings where variable selection is considered, additional steps are needed to derive causal effect estimates. Model building may then entail machine learning and cross-validation \cite{Dukes2020DoublyRobust}.

In this paper we have dissected opportunities and pitfalls arising when drawing real world evidence on the effect of point exposures on survival from incident disease registers. The nature of these registers enables avoidance of immortal time bias through trial emulation. They can be rich in baseline covariate information, especially when linked with additional registers, but typically lack regular time-varying covariates. Long term follow-up of the end point may be needed to inform on relevant patient horizons following treatment decisions. This often comes, however,  with a moving target of  patient cohorts entering over calendar time. It qualifies attainable estimands which must describe their study population well to allow for transportability with and without extrapolation into the future. It also impacts on assumptions and hence on the choice of causal inference for (asymptotically) unbiased estimates of well chosen estimands. We found IP-weighted Kaplan-Meier curves producing biased comparisons due to informative censoring. Standardization through outcome regression avoided this and revealed calendar time trends contributing to the informative nature of administrative censoring. The large differences seen in the case study on PKT versus dialysis first at ESRD, comes with a plea to be specific about study and target population(s) when reporting results and conclusions. This is no less important in subject matter journals which all too often remain fuzzy on such critical points.  The relevance is not restricted to long term disease registers, but equally enters when analyzing shorter term survival from cohorts with fast changing populations as in an emerging pandemic.

\section*{Acknowledgements}
We are grateful to the Steering Group of the Swedish Renal Registry and to Susanne Gabara for assisting us with data transfer. The computational resources (Stevin Supercomputer Infrastructure) and services used in this work were provided by the VSC (Flemish Supercomputer Center), funded by Ghent University, FWO and the Flemish Government – department EWI. In particular, we thank Alvaro Garcia and Kenneth Hoste for their help while running the analyses. We are thankful to Linda Nyanchoka and Raphaël Porcher for their comments and suggestions on an earlier draft and to Dries Reynders who made the independent analyses of the data. 

\section*{Author contributions}
COP contributed in the conception and design of the study, conducted and interpreted the analysis and drafted the manuscript. IW, SS and EG contributed in the conception and design of the study, interpretation of findings and revision of the manuscript. All authors approved the final version.

\section*{Funding}
This project has received funding from the European Union’s Horizon 2020 research and innovation programme under the Marie Sklodowska-Curie grant agreement No 676207. Waernbaum was funded by the Swedish Research council, grant No 2016-00703

\section*{Conflict of interest}
The authors declare no potential conflict of interests.

\bibliographystyle{plain}
\bibliography{PKT}

\appendix
\section{Additional figures}\label{App}
\begin{figure}
    \centering
    \includegraphics[width=0.9\textwidth]{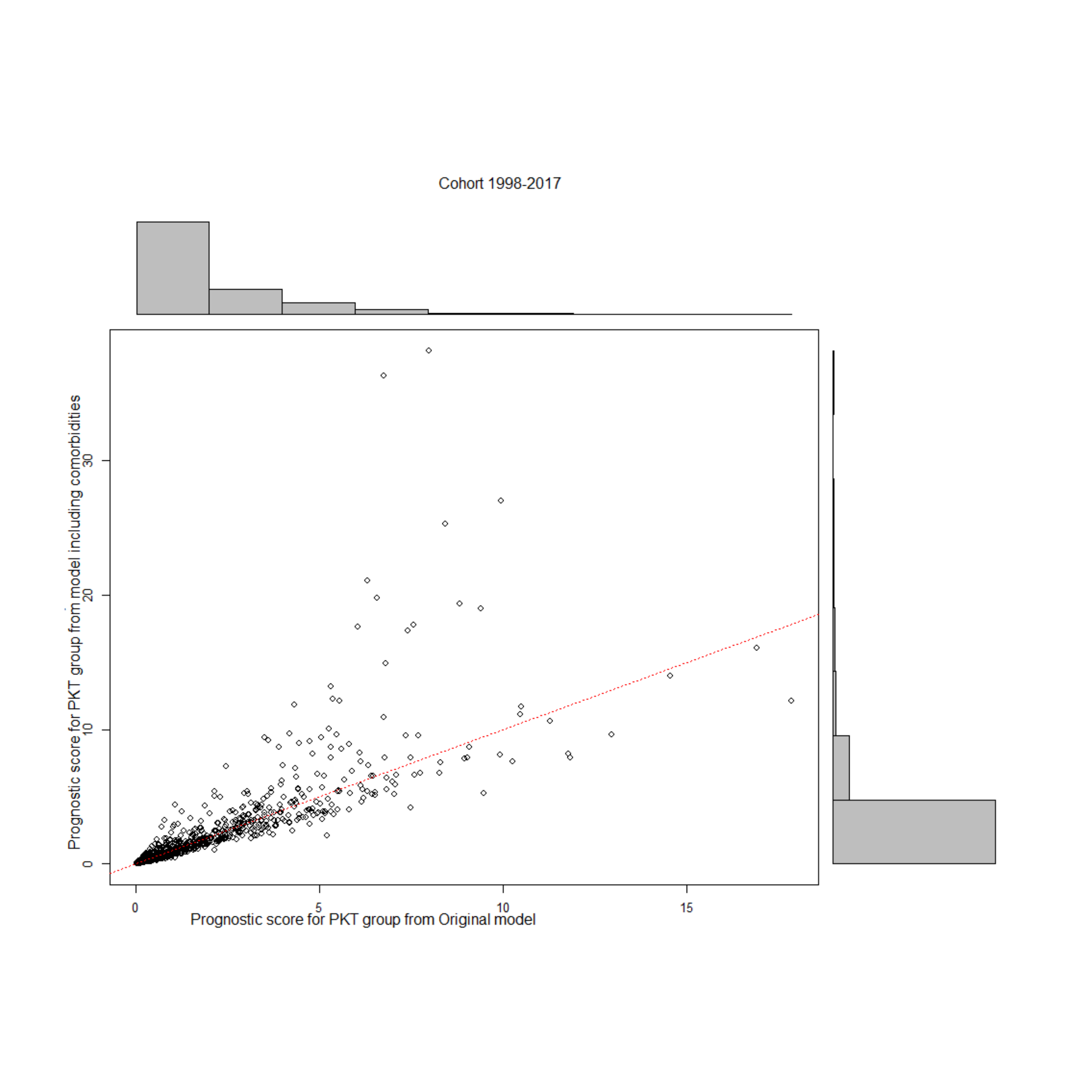}
    \caption{Scatterplot with prognostic scores of the PKT patients as estimated from model B, which includes comorbidities, versus the original model A}
    \label{fig:Scatterplot}
\end{figure}

\begin{figure}
    \centering
    \includegraphics[width=0.9\textwidth]{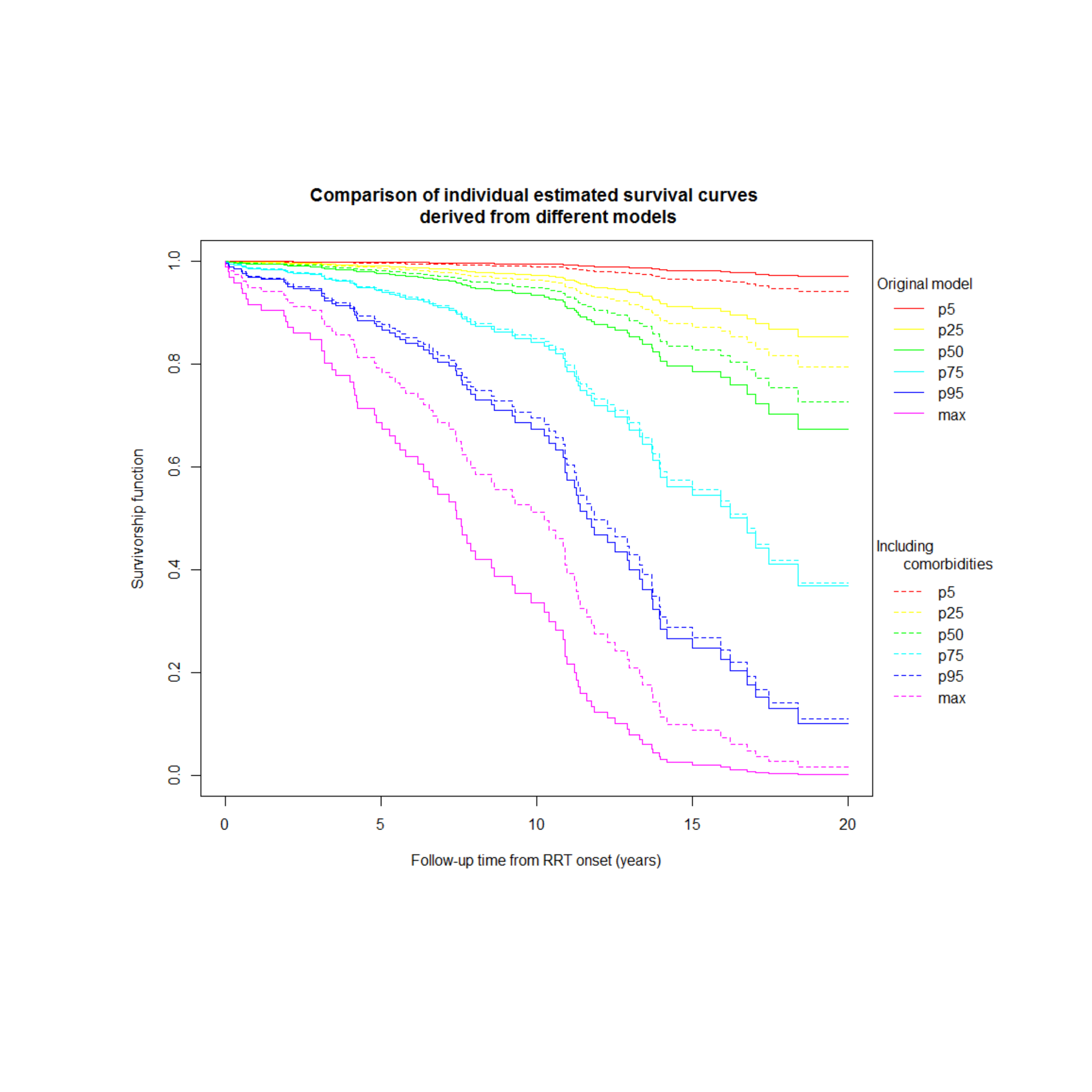}
    \caption{Estimated survival curves derived from the original model A (solid line) compared to model B that adds comorbidities (dashed line) for risk profiles at  p5, p25, p50, p75, p95 and the maximum prognostic score for each model for the cohort 1998-2017}
    \label{fig:Individual}
\end{figure}

\begin{figure}
    \centering
    \includegraphics[width=0.9\textwidth]{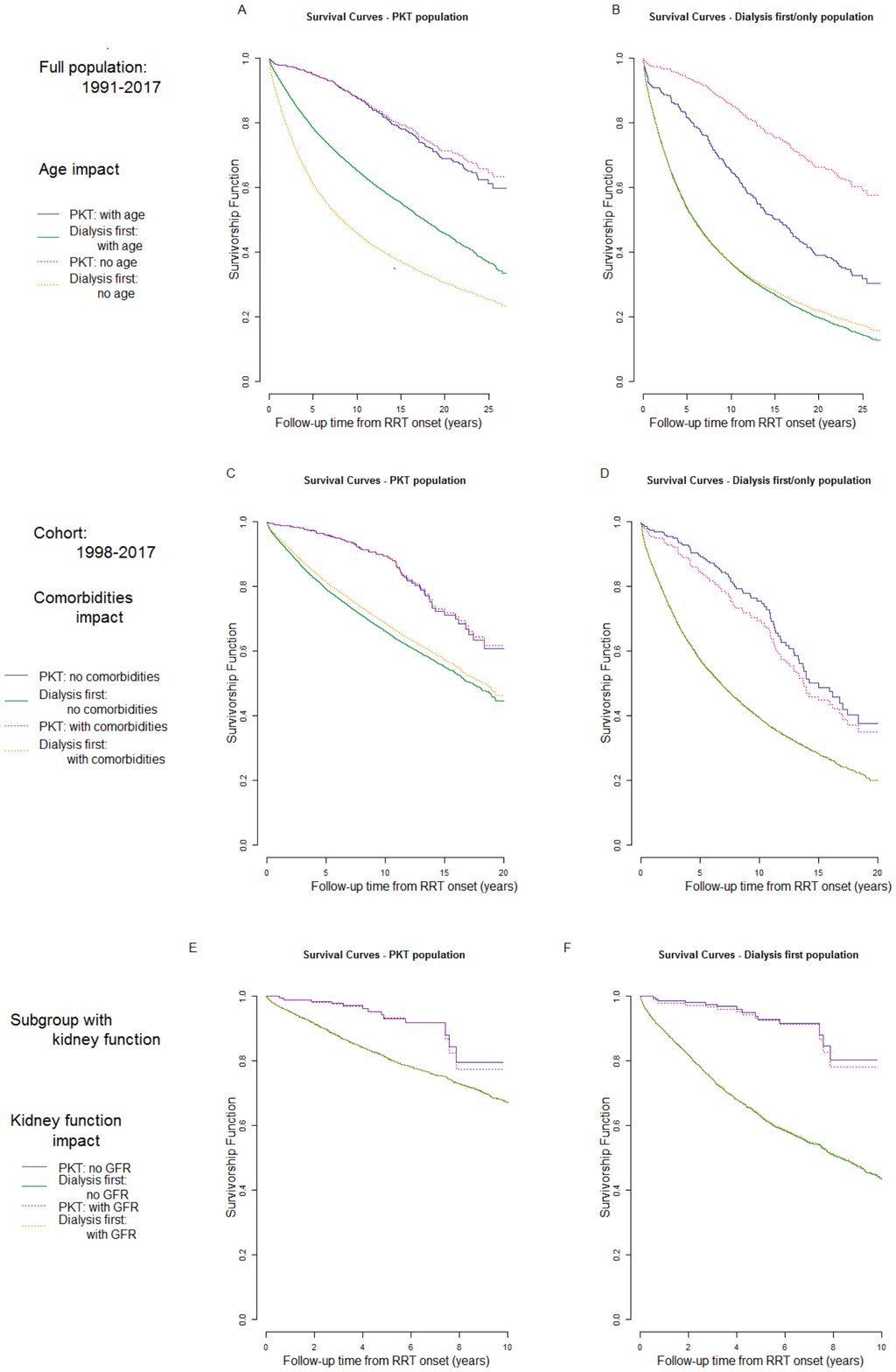}
    \caption{Visualizing the impact on standardized survival of including additional covariates}
    \label{fig:Covariate_impact}
\end{figure}

Below we analyse the subcohort of patients starting RRT from 1998 onwards to illustrate how control for comorbidity indicators affects covariate-specific survival while it leaves the population average survival virtually unchanged. Figure \ref{fig:Scatterplot} plots individual prognostic scores in the PKT group based on  model B (with indicators for diabetes, hypertension and cardiovascular disease in addition to the original covariates age, sex, kidney disease and calendar year), versus model A with just the original covariate set. Figure \ref{fig:Individual} shows (Cox model) derived  survival curves  at observed percentiles p5, p25, p50, p75, p95 and the maximum. For low risk profiles p5 and p25, reduced model A overestimates survival compared to model B. The opposite happens for higher risk profiles p50 and beyond, where the original model underestimates survival. Ignoring comorbidities thus shows substantial impact on estimated covariate specific survival curves, with negligible effect on the population average survival in  the  PKT population (Figure \ref{fig:Covariate_impact}).

\end{document}